\documentclass[USenglish]{article}
\usepackage[utf8]{inputenc}
\usepackage[big,online]{dgruyter}
\usepackage{lmodern} 
\usepackage{microtype}
\usepackage[numbers,square,sort&compress]{natbib}
\usepackage{xcolor, graphicx}
\usepackage{caption, subcaption} 
\usepackage{multirow}
\usepackage{multicol}

\theoremstyle{dgthm}

\theoremstyle{dgdef}

\begin{document}
	\articletype{Research Article}
	\received{Month	DD, 2022}
	\revised{Month	DD, YYYY}
  \accepted{Month	DD, YYYY}
  \journalname{De~Gruyter~Journal}
  \journalyear{YYYY}
  \journalvolume{XX}
  \journalissue{X}
  \startpage{1}
  \aop
  \DOI{10.1515/sample-YYYY-XXXX}

\title{Bayesian estimation of in-game home team win probability for Division-I FBS college football}
\runningtitle{College football win probability model}

\author*[1]{Jason T.~Maddox}
\author[2]{Ryan Sides}
\author[3]{Jane L.~Harvill} 
\runningauthor{Maddox, et al.}
\affil[1]{\protect\raggedright
Syracuse University, Sport Management, Syracuse, NY, U.S., e-mail: jtmaddox@syr.edu}
\affil[2]{\protect\raggedright 
Texas Woman's University, Mathematics and Computer Science, Denton, TX, U.S., e-mail: rsides@twu.edu}
\affil[3]{\protect\raggedright 
Baylor University, Statistical Science, Waco, TX, U.S., e-mail: jane\_harvill@baylor.edu}
	
	
\abstract{\citet{maddoxetal_cbb_2022,maddoxetal_nba_2022} establish Bayesian methods for estimating home-team in-game win probability for college and NBA basketball.  This paper introduces a Bayesian approach for estimating in-game home-team win probability for Division-I FBS college (American) football that uses expected number of remaining possessions and expected score as two predictors.  Models for estimating these are presented and compared.  These, along with other predictors are introduced into two Bayesian approaches for the final estimate of in-game home-team win probability.  To illustrate utility, methods are applied to the 2021 Big XII Conference Football Championship game between Baylor and Oklahoma State.}

\keywords{In-game probability, Pregame probability, Probability estimation, Maximum likelihood, Bayesian estimation, Dynamic prior, Xgboost, Random forest.}

\maketitle

\section{Introduction}
\label{sec:cfb_intro}
Sports analytics has become a well-established area of research.  Work spans more than 30 years and a large range of difficulty.  Since the early 2000s, research in statistical methods for sports analytics has risen dramatically.  The review articles of~\citet{Kubatko_etal_2007}, \citet{fernandez_2019}, and~\citet{terner_franks_2020} provide a fairly comprehensive review for sports analytics for a wide variety of sports, including football.  One problem of interest is predicting the probability that the home team wins during the course of the game, or predicting ``in-game win probability.'' 

Speaking broadly, models for predicting the outcome of a sporting event can be classified based on two objectives: (1) pregame prediction or (2) in-game, or in-play, prediction. Pregame prediction involves determining the outcome of a game before play begins.  Once play begins, the process of predicting the outcome ends.  In contrast, in-game prediction attempts to use the progress during a game to determine win probabilities that vary as a function of in-game variables, for example, elapsed game time or score difference.  The focus of this paper is in-game prediction for college football. 

One early paper on predicting in-game win probability for baseball is attributed to \citet{lindsey_1963}, who used a conditional maximum likelihood estimator to determine the probability the home team wins given the current inning the game and the home team's lead.  \citeauthor{lindsey_1963} used historical information to determine percentages the home team had won in a variety of scenarios, and applied those to the current game to determine current team's estimated win probability as the game progressed.  Up until this work, baseball decisions were based on what would maximize the scoring output for an inning.  In contrast, \citeauthor{lindsey_1963}'s groundbreaking research focused on determining how each decision would affect the probability that a team wins instead of only the change in expected score.  A more recent development in estimating in-game win probability was approached by \citet{benz2019} who modeled college basketball in-game win probability by using a series of logistic regressions at different times throughout games on score differential and pregame win probability. He then smooths the multiple logistic regression models into a single smooth function. \citet{maddoxetal_nba_2022} propose a Bayesian model for the National Basketball Association (NBA) based on time and score differential as the two predictors. Much of their methodology can be extended to college football, but different predictors must be considered. In football, score and time are not as informative for estimating win probability.  A simple consideration of the two sports makes clear a different model is required for each.  For example, in basketball, teams score often and quickly -- often within seconds of each other -- and one possession can result in zero up to five points.  However, it is unusual for a team to score in seconds in a football game, and the possible values for scores are not sequential integers.  Additionally, in football, many other variables contribute to the likelihood of a score, such as field position or down-and-distance.  

Only a few attempts have been proposed for in-game prediction of win probability for (American) football. \citet{lock_nettleton_2014} use a random forest of regression trees with up to ten predictor variables to combine pre-play variables to estimate in-game win probability before any play of an National Football League (NFL) game.
\citet{sports_reference_2012} create a quasi ``black box'' model, where they go into some detail about creating their win probability model using expected points along with pregame win probability and the known standard deviation of the end of the game score differential. However, many of the details used for their model are not provided.  Later, \citet{ruscio_brady_2021} compare the performance of the random forest model by \citeauthor{lock_nettleton_2014} and the model put forth by \citeauthor{sports_reference_2012} when applied to the NFL. Their findings were that there was no discernable difference between the two models. \citeauthor{ruscio_brady_2021} were able to obtain the \citeauthor{sports_reference_2012} model for their paper to reproduce the results.  For Australian rules football, \citet{ryall-2011} used play-by-play data with pregame Elo rankings to develop a model for Australian Rules football.  
In what follows, a new approach that uses Bayesian methods is proposed.  Explanatory variables for predicting in-game win probability include expected number of possessions remaining and expected score differential and with pregame power rankings.

The remainder of the paper is organized as follows. Section~\ref{sec:cfb_data} details the process of gathering and cleaning the data. Section~\ref{sec:cfb_poss} presents a method for modeling the expected number of possession remaining in a college football game.
In Section~\ref{sec:cfb_point}, multiple models are proposed for best estimation of the expected score and compared to each other.  The resulting estimates of expected remaining possessions and expected score are used as predictors for the overall win probability model described in Section~\ref{sec:cfb_winmod}.  The models for win probability are evaluated and applied to a specific game in Section~\ref{sec:cfb_eval}.  Section~\ref{sec:cfb_conclusion} provides closing remarks.

\section{Data Collection}
\label{sec:cfb_data}
The primary goal of the proposed models is effective practical  prediction of in-game home team win probability for a  single college football game or a collection of college football games during the regular season.  The data collected for investigating the proposed models performances were taken from ESPN.  Specifically, play-by-play data from ESPN was scraped using \emph{R} \citep{R} and the package {\tt rvest} \citep{rvest}. For football games, ESPN does not display all plays on each game's webpage by default. Instead, they store the data in a back end server that can be accessed through the ESPN Developer Center. \footnote{The description of the ESPN Developer Center may be found at \url{www.espn.com/apis/devcenter/overview.html}.} There is a general form of the URL for college football back-end data providing access to the play-by-plays.  For each game, the game id changes in the URL. The game ids are scraped for all games in each season, then input into this URL to scrape the play-by-plays for each game. The data was collected starting with the beginning of the 2004 college football season through the 2021 season, excluding the 2020 season due to the uniqueness of that season resulting from the COVID-19 pandemic. 

There were some issues that occurred working with the raw data from ESPN's back-end server. As when working with any raw data, there can be typographical errors or mislabels that must be evaluated and corrected, if correction is possible.  Clear identifier were not always reliable; for example, errors were found on which team had possession of the ball.  The data was combed through many times to ensure that specific identifiers for the game state were correctly labeled.  
For some games, play-by-play data was not available in any capacity.  This was true especially for games during or close to the 2004 season. There was no clear pattern to which games had no play-by-play data.  However, enough games were collected on all season that the missing games should not have a negative impact on the analysis.

\section{Possessions Remaining Model}
\label{sec:cfb_poss}
Within the game of college football, pace or tempo of play has been a key discussion point for the last 20 years.  Mike Leach is considered a modern day football pioneer.  While he was the head football coach at Texas Tech University, Leach ushered in a new offensive play style which he carried with him to Washington State University and then Mississippi State University.   The style of play is commonly called the ``air-raid'' or ``up-tempo'' offense.  A primary feature of the play style is that as little time as possible is used between successive offensive plays.  The philosophy is that minimizing time between snaps prevents the opponent from successfully setting up their defense. 

The most common measure of a team's pace is the team's average number of plays per game. However, \citet{troch_2016} introduces time between consecutive offensive plays as a viable alternative. He argues that average number of plays per game does not take into account the tempo of the opponent or the number of run versus pass plays a team calls.  Because of incomplete passes, pass plays will stop the clock more frequently than run plays.  Therefore, the more pass plays a team attempts, the more plays that team will run during a game, without it necessarily being due to that team's tempo. 

For the purposes of the win probability model, interest lies more in the number of possessions that remain in the game than the time between consecutive offensive plays. The more possessions that remain, the more snaps, and the more potential points are left for teams to score in the game. \citeauthor{troch_2016}'s critique of plays per game not accounting for the opponent is valid.  To account for this, a new method is proposed that is similar to the method put forth by \citet{pomeroy_2012} for college basketball. He calculates pace based on the number of possessions a team would expect to have in a game against a team that plays with average tempo.

In what follows, the term ``pace'' is defined as the expected number of possessions against an opponent that plays with average tempo.  Pace is calculated by a recursive algorithm as described with the following steps for each season.  FAT iteration $m$, for team $k = 1, 2, \ldots, n$, let $\xi_{k, m}$ represent pace.  For a given season, $n$ is the number of Football Bowl Subdivision (FBS) teams.  
\begin{enumerate}
\item For all teams $k = 1, \ldots, n$, initialize $\xi_{k, 0} \equiv 0$. 
\item Choose $\delta > 0$ to be small.\footnote{The value $\delta = 0.0001$ was used.} 
While $\max_k\{|\xi_{k, m} - \xi_{k, m-1}|\}  >  \delta$, 
\begin{enumerate}
    \item Calculate the average pace for all teams,
    \begin{align*}
    \mu_m = \frac{1}{n} \sum_{k = 1}^n \xi_{k, m - 1}.
    \end{align*}
    \item For each team $k = 1, \ldots, n$, let $\kappa_k$ represent the collection of indices corresponding to opponent teams played during the regular season.  For each team $k$, sum the paces of the opponent teams,
    \begin{align*}
    \psi_{k, m} = \sum_{j \in \kappa_k} \xi_{j, m - 1}, \quad k = 1, \ldots, n.
    \end{align*}
    \item Find the average difference between the total number of possessions played $x_{k,m}$ and the expected possessions played $\psi_{k,m}$,
    \begin{align*}
            \epsilon_{k, m} = \frac{x_{k, m} - \psi_{k, m}}{w_k}, \quad k = 1, \ldots, n,
    \end{align*}
    where $w_k$ is the number of games the team $k$ played in the given season.
    \item Assign the updated pace for teams based on their number of possessions above expected,
    \begin{align*}
        \xi_{k, m} = \mu_m + \epsilon_{k, m}, \quad k = 1, \ldots, n.
    \end{align*}
    \end{enumerate}
\item When $\max_k\{|\xi_{k,m} - \xi_{k,m-1}|\} \le \delta$, the convergence criteria is satisfied and the pace $\xi_k$ for each team is taken to be the final iterative value of $\xi_{k. m}$.
\end{enumerate}

The fastest and slowest team paces for Power 5 FBS teams for the 2021 season
are shown in Table~\ref{tab:cfb_pace}. The Power 5 teams with the highest pace are teams that are known for having up-tempo, pass heavy offenses.  For example, Pitt and Tennessee had two of the highest scoring offenses in the nation due to their fast pace. On the other end, teams known for their slow-it-down, run-based offensive style have the lower pace. Kansas State and Boston College are known for their grind-it-out style that does not translate into many possessions or points.
\begin{table}[!ht]
    \caption{Fastest and slowest paces for Power 5 teams for the 2021 regular season.}
    \label{tab:cfb_pace}
    \begin{tabular}{rrc}
        Rank & Team & Pace ($\xi_{k}$) \\ \midrule
        11 & Oklahoma State Cowboys & 30.11 \\
        16 & Duke Blue Devils & 29.69 \\
        21 & Pitt Panthers & 29.13 \\
        22 & Colorado Buffaloes & 29.04 \\
        23 & Tennessee Volunteers & 28.96 \\ \midrule
        124 & Boston College Eagles & 23.96 \\
        125 & Kentucky Wildcats & 23.76 \\
        127 & Oregon Ducks & 23.19 \\
        128 & BYU Cougars & 22.90 \\
        129 & Kansas State Wildcats & 22.42 \\
    \end{tabular}
\end{table}

To compute the value of expected possessions remaining after a play, denoted $\tau$, the two teams' paces are averaged and the average is weighted with a decreasing linear function of the time left in the game.  If $\xi_1$ and $\xi_2$ represent the pace for the two teams in the game, and $t$ the number of seconds elapsed during the game, then $\tau$ is given by
\begin{equation}
\label{eq:tau}
    \tau = \left(\frac{3600 - t}{3600}\right) \left(\frac{\xi_1 + \xi_2}{2}\right).
\end{equation}

\section{Point Value Model}
\label{sec:cfb_point}
There have been many attempts to accurately predict the point value of a drive in football.  Many models are explained in generalities, but the details are vague.  For example, one of the more notable expected point models was posted by a user named ``Mike'' and is featured on  \texttt{Sports-Reference.com} \cite{sports_reference_2012}.  A linear model is used to predict the average point value for a possession given down, distance, and yards to end zone. However the website does not disclose the specifics of the model. The author of this paper contacted  \texttt{SportsReference.com}.  They did not wish to provide any details regarding the model.

Many of the models predict the point value of the current drive.  However, they fail to account for the dependency of the success of the next drive on the outcome of the current drive.
Consider, for example, the situation when a team has a fourth down 99 yards away from the end zone. That team will almost certainly punt, which will typically result in the opposing team receiving the ball with good field position, making that opponent more likely to score.  On the other hand, if a team has the ball just a few yards from the end zone, they are likely to score.  They will then kick the ball off and the opponent would likely start their following possession about 75 yards from the end zone. 
These two contrasting scenarios illustrate the effect of a prior possession on the likelihood of the subsequent possession resulting in a score and thus on the in-game win probability.
Therefore, instead of modeling the point value of only the current possession, the proposed model is built on the combined score for both the current possession and the following opponent possession.

Four competing models are used to predict the point value of the current and ensuing possessions using the predictor variables in Table~\ref{tab:predvars}.  The first model is a linear regression model with no interactions.  The second is a linear regression that includes all $2^6 = 64$ interactions between the nine predictors, with the exception of any interactions involving offensive pace, defensive pace, or number of possessions.  The third and fourth models are a random forest model and an extreme gradient boosting (XGBoost) model, respectively.  Literature on linear regression is prevalent, and the technique is commonly applied.  However, random forests and the XGBoost models are more current, lesser known techniques.  Random forest and XGBoost models have become more important with advances in computation.  Sections~\ref{sec:cfb_random_forest} and~\ref{sec:cfb_XGBoost} provide overviews of these techniques.  For more detail on random forest see~\citet{James2013}, and for more detail on XGBoost, see~\citet{chen_guestrin_2016} and~\citet{jain_2022}.


\begin{table}[!ht]
    \caption{Predictor variables in competing models for predicting the point values of the current and ensuing possessions.}
    \label{tab:predvars}
    \begin{tabular}{lll} 
    Time remaining ($3600 - t$) & Offensive score & Defensive score \\
    Down & Distance & Yards to end zone \\
    Number of possessions played by time $t$ & Offensive pace & Defensive pace \\
    \end{tabular}
\end{table}


\subsection{Random Forest Model}
\label{sec:cfb_random_forest}

The random forest model is built on many independent decision trees. A decision tree is a machine learning algorithm that takes the data as a whole, finds the ``best'' predictor/value of predictor pair to split the data into two groups, or child nodes. What is ``best'' is determined by the predictive power, that is, the one that would minimize root mean square error (RMSE) or mean absolute error (MAE) of the model, using the average response in each node as the prediction for the observations that fall in that node. Each of these child nodes are then split again by the same process, with splitting continuing to occur until some stopping criteria is reached. This stopping criteria could be the maximum number of splits in a tree, the maximum depth of the tree, or the maximum number of terminal nodes.

To build a random forest, a sample with replacement is taken from the data, typically the same size as the data. This sample is used to build a decision tree, then another sample, independent from the previous, is taken and another tree is built. This is repeated many times to gather many independent trees. To make a prediction, an observation is inserted into each tree, with the prediction being the average response from its terminal node. These predictions from each of the trees are then averaged to obtain the overall prediction for the random forest.

Machine learning models, such as random forests, do not require assumptions such as statistical models do. Instead, they have parameters that require ``tuning.'' For example, the maximum number of splits in a tree is a parameter that could be tuned for each random forest model. The process would be to run the model on several different values for the maximum number of splits, then the number that minimizes the MAE of the test data would be chosen for this model. The fewer number of splits in the tree, the less fit each tree becomes to the training data, causing a greater risk of underfitting the data. Likewise, the larger number of splits in each tree, the greater the risk of overfitting the training data. Parameters for random forests can be split into two categories, tree-based parameters and model-based parameters. Tree-based parameters are applied to each tree individually and include, but are not limited to maximum number of splits, minimum number of observations in a terminal node, and maximum parameters considered at each split. Model-based parameters are values pertaining to the random forest as a whole and include, but are not limited to number of trees in the model and probabilities attached to each observation of being included in each sample for building the tree.

\subsection{XGBoost Model}
\label{sec:cfb_XGBoost}
An extreme gradient boosting model is similar to a random forest in that it is a machine learning algorithm that is built on numerous decision trees. The difference lies in that instead of the trees being independent from another, each tree is built one at a time on the errors of the previous tree in XGBoost. Each tree is also weighted down by a shrinkage factor, $\eta,~0 < \eta < 1$, to help prevent the tree from learning too quickly and over-fitting the training data. In the training set, denote ${\mathbf{x}}$ as the collection of all predictor variables in the training set and $x_i$ as the vector of predictor values of observation $i,~y_i$ be the response variable value, and $\hat f(x_i)$ be predicted value from the model $\hat f({\mathbf{x}})$ at $x_i$.
The XGBoost algorithm can be set up as follows:
\begin{enumerate}
    \item Initialize $\hat f^0(x_i) \equiv 0$ and $r_i^0 = y_i$ for all $i$ in the training set, where $r_i^0$ is the residuals for each observation before any iterations in the algorithm.
    \item For $b = 1, 2, \ldots, B$, where $B$ is the total number of trees in the algorithm,
    repeat the following:
    \begin{enumerate}
        \item Fit a tree $\hat f^b(\mathbf{x})$ to $r^{b - 1}$, the residuals from the most recently updated model.
        \item Update $\hat f^b(\mathbf{x})$ by adding a shrunken version of the new tree,
        \begin{align*}
            \hat f(\mathbf{x}) \leftarrow \hat f(\mathbf{x}) + \eta\hat f^b(\mathbf{x}).
        \end{align*}
        \item For each $i$, update the residuals,
        \begin{align*}
            r^b_i \leftarrow r^{b - 1}_i - \eta\hat f^b(x_i).
        \end{align*}
    \end{enumerate}
    \item Output the boosted model,
    \begin{align*}
        \hat f(\mathbf{x}) = \sum_{b = 1}^B\eta\hat f^b(\mathbf{x}).
    \end{align*}
\end{enumerate}

Choice of $B$ is critical, since for larger values of $B$, that is, a model with more trees, $\eta$ must be closer to zero to prevent over-fitting the training data. If $\eta$ is large, say close to 1, then the first tree built will start to fit the training data well, and trees later on in the algorithm will be fitting residuals that are more likely due to noise than to an actual signal.

Tuning of the XGBoost model can often be time consuming due to the number of parameters to tune around and the amount of time it can take to run each iteration of the model for large data sets. XGBoost has the same tree-based parameters that are present in random forest models, most notably the depth of the tree, because the depth of each tree is the largest possible interaction depth the model can detect. The additional model-based parameters that need tuning are $B$ and $\eta$ which are tuned simultaneously. There are many methods for tuning parameters of the XGBoost model, one of which is 
performed using the following steps.
\begin{enumerate}
    \item Choose a relatively high $\eta$, somewhere between 0.05 and 0.2.
    \item Optimize $B$ for this shrinkage value, keeping $B$ to a value where a machine can run the model relatively quickly.
    \item Tune tree-based parameters using the values of $\eta$ and $B$ obtained in the first two steps.
    \item Decrease $\eta$ and increase $B$ proportionally until the model's performance improves minimally.  Measures of the model's performance include statistics like MAE or RMSE.
\end{enumerate}

\subsection{Test and Training Data for Models}
\label{sec:ttdata}
To fit the point value model, and later the win-probability model, the data was separated into a test data set and two training data sets.  In particular, the last five of the 17 seasons were designated as the test data for the win probability model described in Section~\ref{sec:cfb_winmod} and evaluated in Section~\ref{sec:cfb_eval}.  For the remaining 12 seasons, half of the data were randomly selected to act as the overall data for the point value model.  This data will be referred to as the ``point value data.''  The other half were used to build the win probability model.

\subsection{Point Value Model Selection}
\label{sec:ptvaluemodel}
To determine which of the four point value models is best, the point value data was further separated into test and training data sets.  Each of the four point value models were built on the training set that was randomly pulled from the point value data and then applied to the point value test data. The models were compared using mean absolute error (MAE). The results are shown in Table~\ref{tab:cfb_point}. For both linear regression models, the average difference between the predicted point difference for the current and following possession and the actual point difference was slightly more than three points per 
play. The random forest model slightly outperforms the regression models, lowering the MAE to slightly below three points. The XGBoost model performs the best, with the lowest MAE by a margin at least 0.2949 points.  Therefore, the XGBoost model is adopted to predict the point differential for the current and subsequent possessions.  Conclusions about model performance are similar based on root mean square error.

\begin{table}[!ht]
    \caption{Performance of point value models using test MAE.}
    \label{tab:cfb_point}
    \begin{tabular}{rc}
        Model & MAE \\ \midrule
        Linear regression & 3.0805 \\
        Regression w/ inter & 3.0614 \\
        Random forest & 2.9751 \\
        XGBoost & 2.6802 \\
    \end{tabular}
\end{table}

\section{Win Probability Model}
\label{sec:cfb_winmod}
For a specific game, consider the random process that is the expected lead for the home team following the current and succeeding possession, which will now be referred to as expected score, $\omega$. The expected score will shift as time in the game moves forward, or equivalently as the expected possessions remaining in the game, $\tau$, decreases, where $\tau$ is calculated as in~\eqref{eq:tau}.  For specific values of $\tau$ and $\omega$, let $p_{\tau, \omega}$ represent the in-game probability that the home team will win the game.  When considering multiple games $i = 1, 2, \ldots, M$, let $Y_i = 1$ if the home team wins game $i$ and 0 otherwise. Consider $p_{\tau, \omega}$ as a continuous function of $\tau$ and $\omega$. \citet{maddoxetal_cbb_2022}~introduce a Bayesian estimator of in-game win probability, $p_{t, \ell}$, based on current time $t$ and score differential $\ell$, in college basketball.  Their methods are extended and adapted to the NBA in~\citet{maddoxetal_nba_2022}.  In both papers, \citeauthor{maddoxetal_cbb_2022} argue that the nature of a basketball game makes time and score differential excellent predictors for in-game win probability.  For college football, instead of using time and score differential as predictors, $\tau$ and $\omega$ are preferred for two reasons.  First, and most obviously, the nature of the games of basketball and football are inherently different.  As opposed to basketball, the predictors $\tau$ and $\omega$ contain more information about current the state of the game, and account for which team has the ball, how physically close (on the field) that team is to scoring, etc.  We now extend the methods of~\citet{maddoxetal_nba_2022} to estimate $p_{\tau, \omega}$ for college football.

\subsection{Naive Estimator of In-game Win Probability}
\label{sec:cfb_naive}
For each combination of $\tau$ and $\omega$ rounded to the nearest whole number, consider the $(\tau, \omega)$ ``cell.''  On a specific cell, the number of wins by the home team, $n_{\tau, \omega}$, follows a binomial$(N_{\tau, \omega}, p_{\tau, \omega})$ distribution, where $N_{\tau, \omega}$ is the total number of games that are observed in the $(\tau, \omega)$ cell. Due to the total number of possible cells, one for each combination of $\tau$ and $\omega$, some cells may have a small value of $N_{\tau, \omega}$, resulting in a large standard error for any estimator for $p_{\tau, \omega}$. \citet{deshpande_jensen_2016} and \citet{maddoxetal_cbb_2022,maddoxetal_nba_2022} suggest a binning approach to address the small sample sizes. Windows centered around $(\tau, \omega)$ can be created in such a way that the in-game win probability remains relatively constant across the window. For creating the interval around expected score, since scores were considered for each two possessions, a shift of two possessions will rarely have a major affect on win probability.  Moreover within college football for similar score differentials, two or fewer possessions should not have a large effect on the win probability, especially early in the game. Therefore a reasonable interval for expected possessions remaining is defined as $[\tau - 2, \tau + 2]$.  To determine an interval for the expected score, note that when an offense scores, the fewest points attainable is three points from a field goal. Therefore, the interval on expected score is taken to be $[\omega - 3, \omega + 3]$. The same notation will be adopted for any $[\tau - 2, \tau + 2] \times [\omega - 3, \omega + 3]$ window; that is, $N_{\tau, \omega}$ is the number of games in the window in which the home team has an expected lead by any value in $[\omega - 2, \omega + 2]$ points with any expected possessions remaining in $[\tau - 2, \tau + 2]$,  and given the specific value of $N_{\tau, \omega},~n_{\tau, \omega} = \sum_{i=1}^{N_{\tau, \omega}} Y_i$, distributed as a binomial($N_{\tau, \omega}, p_{\tau, \omega})$ random variable. Based on the binomial distribution, a simple estimator for in-game home team win probability for for each $(\tau, \omega)$ window is the maximum likelihood estimator
\begin{equation}
    \label{eq:cfb_MLE}
    \bar p_{\tau, \omega} = \frac{n_{\tau, \omega}}{N_{\tau, \omega}}.
\end{equation}
As a given game approaches the end of regulation, each individual point and possession will have a larger impact on the in-game win probability. Therefore the windows should be modified at the end of the game to reflect this.
Starting from fifteen expected possessions remaining, the proposed method shortens window lengths and widths; that is, the length of interval around each of expected score and expected number of possessions remaining will gradually decrease until there is an expected two possesses remaining, when the intervals' widths become zero.

\subsection{Dynamic Bayesian Estimator}
\label{sec:cfb_dyn_bayes_est}
To elicit a prior distribution for in-game win probability,
\citet{maddoxetal_nba_2022} suggest 
polling a sample of industry experts. The same can be done for college football.  The authors contacted a panel of 14 college football experts, including coaches from a major Division I college football team and respected media personnel.  Each provided their estimate of the probability a team wins for each combination of score differential, $\ell$, and time elapsed, $t$, in Table~\ref{tab:cfb_prior}, regardless of which team is the home team.   Note that $t$ and $\ell$ are the actual time and score differential in the hypothetical game and are not the same as the previously defined $\tau$ and $\omega$.  To reduce the level of complexity and increase the likelihood of response, the authors choose to ask the question of win probability in terms of $t$ and $\ell$ as opposed to $\tau$ and $\omega$.

The Bayesian prior parameters have an interesting interpretation, first noted by \citet{deshpande_jensen_2016}. The parameter $\alpha_{t, \ell}$ can be interpreted as the number of ``pseudo-wins'' in the $(t, \ell)$ cell; likewise $\beta_{t, \ell}$ as the number of ``pseudo-losses.'' Through this interpretation, the two parameters can be seen as a way of increasing the number of games in a specific $(t, \ell)$ cell. If the home team is ahead, then first scale parameter being large effectively acts to increase the number of wins in that cell.  On the other hand, if the home team is behind, the second scale parameter is large, and acts to increase the number of losses.

The sample mean $\tilde p_{t, \ell}$ and sample variance $s^2_{t, \ell}$, of the probabilities were computed. The two scale parameters were estimated via a method-of-moments type approach. The system of equations 
\begin{align*}
    \tilde p_{t, \ell} & = \frac{\alpha_{t, \ell}}{\alpha_{t, \ell} + \beta_{t, \ell}}, \\
    s^2_{t, \ell} & = \frac{\alpha_{t, \ell}\beta_{t, \ell}}{\left(\alpha_{t, \ell} + \beta_{t, \ell}\right)^2\left(\alpha_{t, \ell} + \beta_{t, \ell} + 1\right)},
\end{align*}
is solved for $\alpha_{t, \ell}$ and $\beta_{t, \ell}$, yielding
\begin{align*}
    \alpha_{t, \ell} & = -\frac{\tilde p_{t, \ell}\left(\tilde p_{t, \ell}^2 - \tilde p_{t, \ell} + s_{t, \ell}^2\right)}{s^2_{t, \ell}} \\
    \beta_{t, \ell} & = \frac{\left(\tilde p_{t, \ell} - 1\right)\left(\tilde p_{t, \ell}^2 - \tilde p_{t, \ell} + s^2_{t, \ell}\right)}{s_{t, \ell}^2},
\end{align*}
as long as $\left(\tilde p_{t, \ell} - 1\right)\tilde p_{t, \ell}\left(\tilde p_{t, \ell}^2 - \tilde p_{t, \ell} + s^2_{t, \ell}\right) \neq 0$.

In Table~\ref{tab:cfb_prior} the score differential is presented when the home team has the lead. The larger the lead, the more the prior is left-skewed. On the other hand, if the visiting team has the lead, then the roles of $\alpha_{t, \ell}$ and $\beta_{t, \ell}$ are reversed, causing the prior to be right-skewed. At any time, if the game is tied ($\ell = 0$), or is sufficiently close in score for the amount of time remaining, the prior distribution is a diffuse beta$(1, 1)$ prior, allowing the likelihood to drive the results in the posterior.

\begin{table}[!ht]
    \caption{Imputed parameters for beta prior.}
    \label{tab:cfb_prior}
    \begin{tabular}{llcc}
        Elapsed Time (t) & Home Team & & \\
        (sec.) & Lead ($\ell$) & $\alpha_{t, \ell}$ & $\beta_{t, \ell}$ \\ \midrule
        $[0, 900]$ & $[0, 7]$ & 1 & 1 \\
        $[0, 900]$ & $(7, 14]$ & 21 & 10 \\
        $[0, 900]$ & $(14, 28]$ & 16 & 3 \\
        $[0, 900]$ & $(28, \infty)$ & 59 & 1 \\
        $(900, 1800]$ & $[0, 7]$ & 1 & 1 \\
        $(900, 1800]$ & $(7, 14]$ & 15 & 7 \\
        $(900, 1800]$ & $(14, 28]$ & 11 & 2 \\
        $(900, 1800]$ & $(28, \infty)$ & 44 & 1 \\
        $(1800, 2700]$ & $[0, 7]$ & 1 & 1 \\
        $(1800, 2700]$ & $(7, 14]$ & 14 & 4 \\
        $(1800, 2700]$ & $(14, 28]$ & 13 & 1 \\
        $(1800, 2700]$ & $(28, \infty)$ & 126 & 1 \\
        $(2700, 3300]$ & $[0, 3]$ & 1 & 1 \\
        $(2700, 3300]$ & $(3, 7]$ & 28 & 16 \\
        $(2700, 3300]$ & $(7, 14]$ & 27 & 8 \\
        $(2700, 3300]$ & $(14, 21]$ & 11 & 1 \\
        $(2700, 3300]$ & $(21, \infty)$ & 98 & 1 \\
        $(3300, 3480]$ & 0 & 1 & 1 \\
        $(3300, 3480]$ & $(0, 3]$ & 29 & 18 \\
        $(3300, 3480]$ & $(3, 7]$ & 17 & 6 \\
        $(3300, 3480]$ & $(7, 10]$ & 13 & 2 \\
        $(3300, 3480]$ & $(10, 14]$ & 16 & 1 \\
        $(3300, 3480]$ & $(14, \infty)$ & 98 & 1 \\
        $(3480, 3600)$ & 0 & 1 & 1 \\
        $(3480, 3600)$ & $(0, 3]$ & 21 & 10 \\
        $(3480, 3600)$ & $(3, 7]$ & 18 & 5 \\
        $(3480, 3600)$ & $(7, 10]$ & 16 & 1 \\
        $(3480, 3600)$ & $(10, \infty)$ & 91 & 1 \\
    \end{tabular}
\end{table}

The maximum likelihood estimator in~\eqref{eq:cfb_MLE} is based on $\tau$ and $\omega$.  
However, the elicited prior is based on $t$ and $\ell$.  
Consequently, the in-game win probability
will be written $p_{t, \ell, \tau, \omega}$.  Since the beta family of distributions is a conjugate prior for the binomial distribution, the beta-binomial connection is used to estimate $p_{t, \ell, \tau, \omega}$ with the mean of the posterior beta distribution, specifically
\begin{equation}
    \label{eq:cfb_posterior}
    \hat p_{t, \ell, \tau, \omega} = \frac{n_{\tau, \omega} + \alpha_{t, \ell}}{N_{\tau, \omega} + \alpha_{t, \ell} + \beta_{t, \ell}}.
\end{equation}

\subsection{Adjusted Dynamic Bayesian Estimator}
\label{sec:adjdynbayesest}
During a game, the in-game win probability is clearly a function of expected possessions remaining and expected score.  However, it is also affected by the overall skill of the teams playing the game.  The skill level can be incorporated using pregame win probabilities for each game. The normal distribution quantile function was used to convert the pregame point spread to a pregame home team win probability $\hat p_p$ for each game using the method outlined by \citet{maddoxetal_cbb_2022}. 

\citet{maddoxetal_nba_2022} introduce three different functions to incorporate pregame probabilities. The first is a linear function of only time remaining. The second is a linear function of time remaining and score differential. The third is linear in time, but quadratic in score differential. Other variations to the three weight functions were considered. For example, including a quadratic term for time was attempted. However, the more complicated models would not converge appropriately for \emph{R} to optimize the performance accurately.
The three weight functions specifically considered here are
\begin{align*}
    D_1 & = bt, \\
    D_2 & = c_0 + c_1t + c_2|\ell|, \\
    D_3 & = d_0 + d_1t + d_2|\ell| + d_3\ell^2.
\end{align*}
Each of the weight functions yields a competing model for including pregame win probabilities, which can be represented as
\begin{equation*}
    p^*_{t, \ell, \tau, \omega, j} = \begin{cases} \hat p_p, & D_j \leq 0 \\ \left(1 - D_j\right) \hat p_p + D_j\hat p_{t, \ell, \tau, \omega}, & 0 < D_j < 1 \\ \hat p_{t, \ell, \tau, \omega}, & D_j \geq 1\end{cases}, \qquad j = 1, 2, 3,
\end{equation*}
where $p^*_{t, \ell, \tau, \omega, j}$ is the final predicted win probability associated with weight function $D_j$. 

\begin{figure}[!ht]
\caption{Graphical function of $D_2$.}
\label{fig:cfb_pct_in_game}
\includegraphics[width=0.8\textwidth]{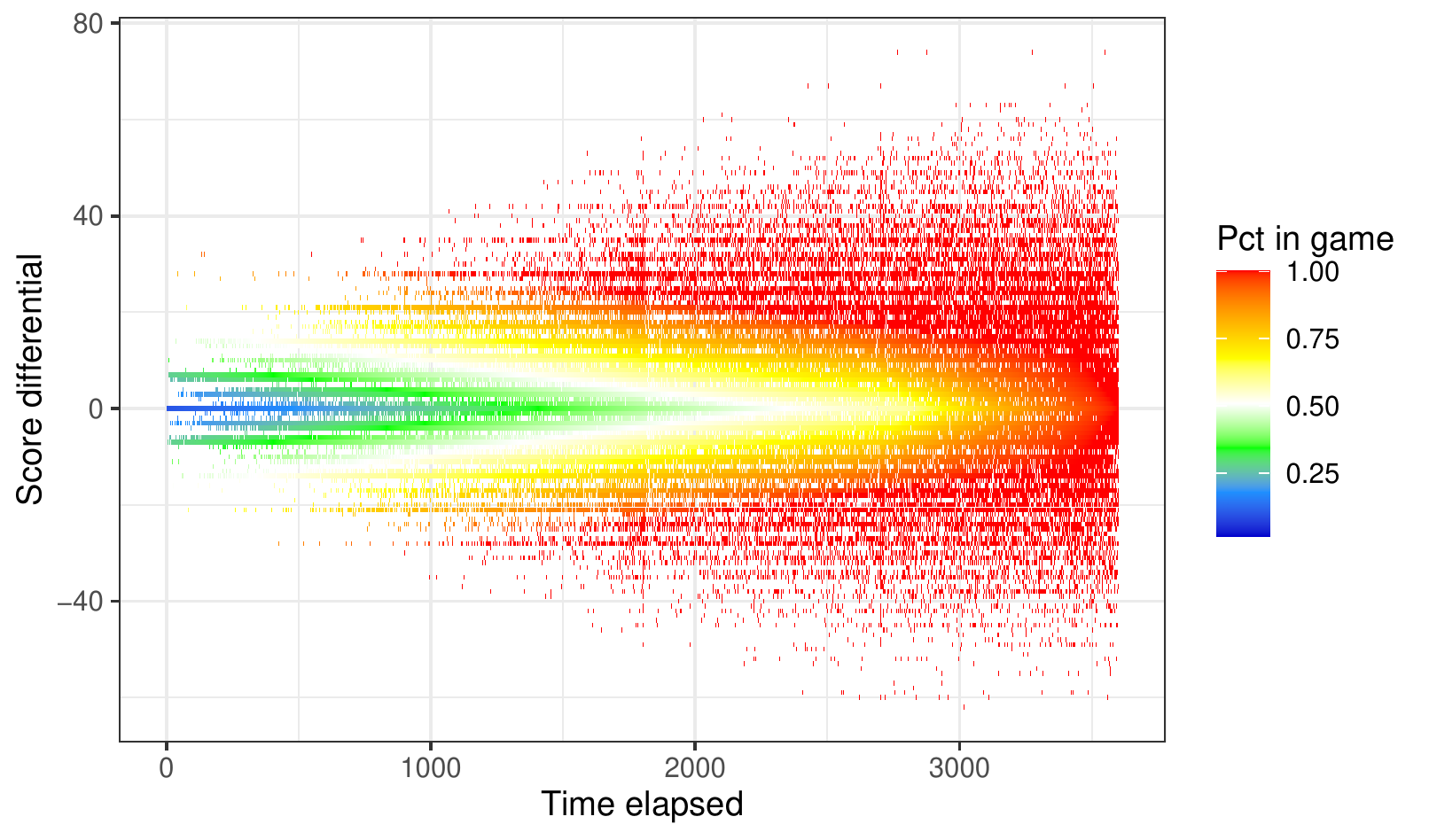}
\end{figure}

Values for $b, c_0, \ldots, d_3$ are estimated by minimizing Brier score for each $p^*_{t, \ell, \tau, \omega, j}$ computed from the holdout test data. The Brier scores for each $p^*_{t, \ell, \tau, \omega, j}$ are shown in Table~\ref{tab:cfb_pct_in_game}. The model $p^*_{t, \ell, \tau, \omega, 2}$ is the most accurate predictor for the test data. The fitted model for $D_2$ that provided the minimal Brier score is
\begin{align*}
    D_2 = 0.09589 + 0.00018 t + 0.02523 \ell.
\end{align*}
Using this expression for $B_2$, the model
\begin{align*}
    p^*_{t, \ell, \tau, \omega, 2} = \begin{cases}\hat p_p, & D_2 \leq 0 \\ \left(1 - D_2\right) \hat p_p + D_2\hat p_{t, \ell, \tau, \omega}, & 0 < D_2 < 1 \\ \hat p_{t, \ell, \tau, \omega}, & D_2 \geq 1\end{cases}
\end{align*}
is the ``adjusted dynamic Bayesian estimator.'' The percent of in-game win probability accounted for in the final predicted win probability is displayed in Figure~\ref{fig:cfb_pct_in_game}.  As desired, early on in any game the majority of the final predicted win probability comes from the pregame win probability. However, as the game draws to an end or one team takes a large lead, the final predicted win probability gets closer and closer to the dynamic Bayesian estimator.

\begin{table}[!ht]
    \caption{Brier scores for models determining in-game probability proportion.}
    \label{tab:cfb_pct_in_game}
    \begin{tabular}{rc}
        Proportion Model & Brier Score \\ \midrule
        Linear Time ($D_1$) & 0.1272 \\
        Linear Time \& Score ($D_2$) & 0.1250 \\
        Quadratic ($D_3$) & 0.1265 \\
    \end{tabular}
\end{table}

\section{Model Evaluation and Application}
\label{sec:cfb_eval}

For each play from each game from the 2017 through 2021 seasons, excluding the 2020 season, Brier score was computed. The Brier scores of the three models, dynamic Bayesian estimator, adjusted dynamic Bayesian estimator, and the random forest model proposed by \citet{lock_nettleton_2014} are shown in Table~\ref{tab:cfb_perform}.  
Brier's score is a statistic used to compare the performance of different methods for estimating probabilities.  Brier's score is the average of the square of the difference between the estimated probability and the observed binary outcome.  In the context of in-game home team win probabilities, this observed binary is $Y_i$ as defined in Section~\ref{sec:cfb_winmod}.  To interpret Brier's score, if $Y_i = 1$ for all $i$, and the predicted probability is also one for every $i$, then Brier's score will be zero, indicating perfect prediction.  On the other hand, if for all $i, Y_i = 0$, and the estimated probability is one, then Brier's score will be one, the worst possible Brier's score.  Both Bayesian models outperform the random forest model. The dynamic Bayesian model with the pregame adjustment is the model that performs best overall.

\begin{table}[!ht]
    \caption{Brier scores for predictive performances for 2017 through 2021 seasons.}
    \label{tab:cfb_perform}
    \begin{tabular}{lc}
        Model & Brier Score \\ \midrule
        Dynamic Bayes & 0.1453 \\
        Adjusted dynamic Bayes & 0.1250 \\
        Random forest & 0.1705 \\
    \end{tabular}
\end{table}

\begin{figure}[!ht]
\caption{In game win probability for 2021 Big 12 Championship.}
\label{fig:cfb_game}
\includegraphics[width=0.8\textwidth]{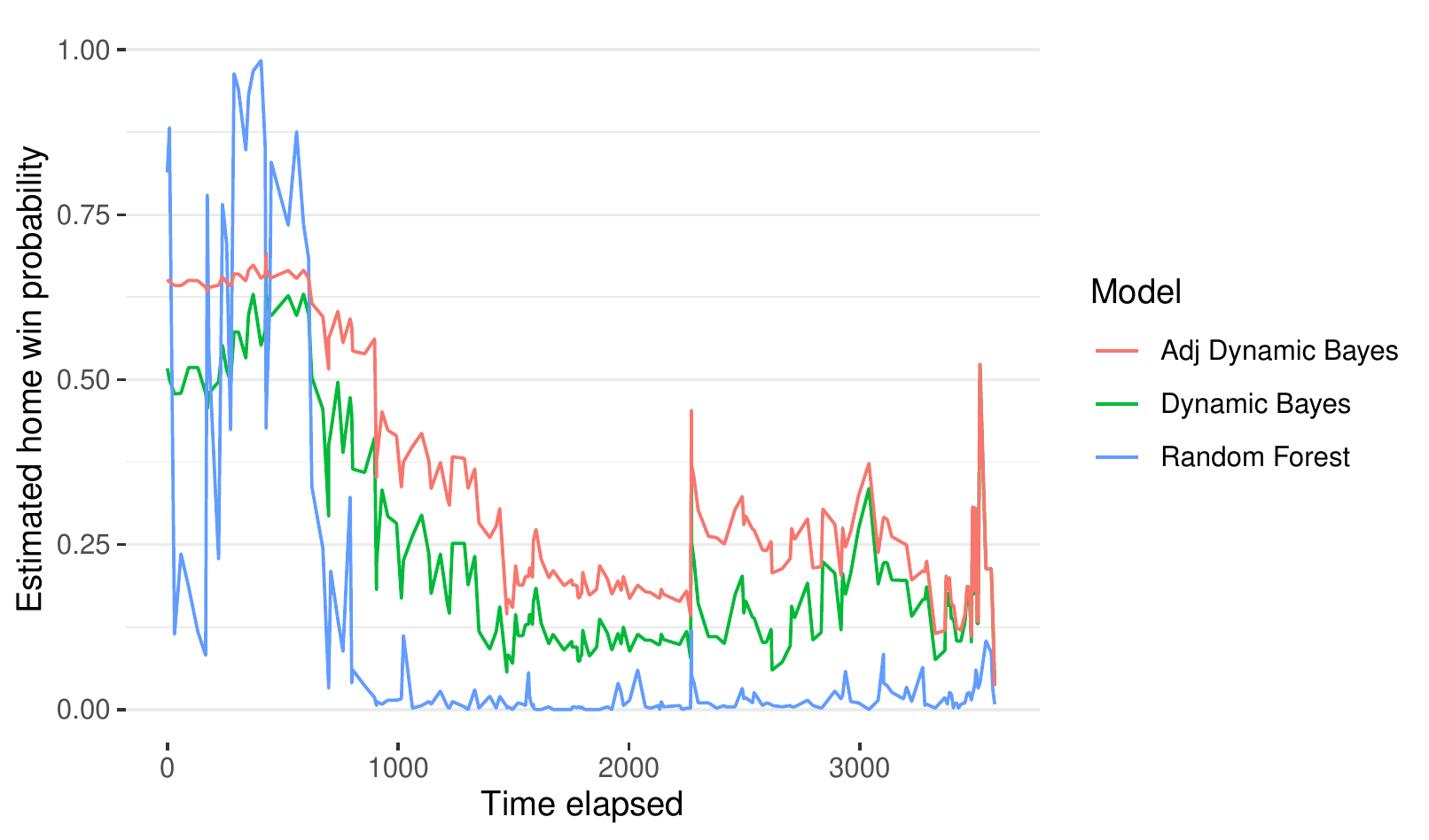}
\end{figure}

To observe the performance of each model further, each model can be applied to a specific game, observing the features of the models as the game progresses. The models are run for the 2021 Big 12 Championship game between Baylor University and Oklahoma State University. The results can be seen in Figure~\ref{fig:cfb_game}.

On December 4, 2021, the Baylor Bears and Oklahoma State Cowboys met at AT\&T Stadium in Dallas, Texas for the Big 12 Championship game.  Oklahoma State was the home team.  During the regular season, Oklahoma State had beaten Baylor, and were considered the favorite to win this game.  The probability traces for the game for each of the three models is seen in Figure~\ref{fig:cfb_game}. The adjusted dynamic Bayes model (red curve) shows that Oklahoma State is favored by starting at well over 50\% chance that Oklahoma State would win.  Early in the game, one of the features that leads to the Bayesian models outperforming the random forest (blue line) is can be seen.  The random forest model is too quick to jump to a win probability close to 0 or 1, especially early in the game.  The game got off to a relatively slow start, but by halftime Baylor had surprised many by jumping out to a 21-6 lead.  At that time, all three models predicted Baylor was most likely to win.  In the second half, Oklahoma State started to make a come back. Halfway through the third quarter, Oklahoma State was able to score a touchdown and make the score 21-13, significantly raising their probability of winning. The Cowboys then had another significant uptick in their win probability with ten minutes to go when they earned a first-and-goal from the Baylor one-yard line, appearing to be on the verge of scoring a touchdown with the potential to tie the game.  However, Baylor was able to hold Oklahoma State to just a field goal, maintaining their lead and their edge in win probability. The climax of the game came with three minutes left when Oklahoma State started a drive from their own 10 yard line. They methodically drove the ball down the field and once again ended up with a first-and-goal, this time from the two-yard line with 80 seconds left. Being down by 5 points, they were not going to kick a field goal.  Instead, they had four attempts to score a touchdown to win the game. With so many attempts from so close, it appeared likely that Oklahoma State would be able to score, driving their win probability in the two Bayesian models over 50\% despite trailing. Once again the Baylor defense made an impressive goal line stand, keeping Oklahoma State out of the end zone, ultimately less than half of a yard short, sealing the victory for Baylor.

\section{Conclusion}
\label{sec:cfb_conclusion}
Two new methods are proposed for estimating in-game win probability for college football games. Both are an extension and enhancement of the methods in~\citet{maddoxetal_cbb_2022,maddoxetal_nba_2022}, which provide a number of models for estimating in-game home-team win probabilities for NCAA basketball and NBA basketball respectively.  The first proposed ``dynamic Bayesian estimator,'' uses expected score differential and expected possessions remaining as predictors, which are modeled from the data themselves, and a prior that has been calculated based on the distribution of predicted win probabilities from 14 college football field experts. The second method, referred to as the ``adjusted dynamic Bayesian estimator,'' adjusts the dynamic Bayesian estimator based on pregame win probabilities obtained from \url{TeamRankings.com}. The adjustment is optimized over a function of both time and score so that as the game moves on or the score differential increases, the adjusted dynamic Bayesian estimator will begin to approach the dynamic Bayesian estimator rather than the pregame win probability. These two methods are then compared to the random forest win probability model proposed by \citet{lock_nettleton_2014}. Both new models outperform the standard random forest model, with the adjusted dynamic Bayesian estimator performing the best out of all of the models.

\bibliographystyle{apalike}
\bibliography{refs}

\begin{thebibliography}{}

\bibitem[Benz, 2019]{benz2019}
Benz, L. (2019).
\newblock A new ncaahoopr win probability model.
\newblock [Online; posted December 26, 2019].

\bibitem[Chen and Guestrin, 2016]{chen_guestrin_2016}
Chen, T. and Guestrin, C. (2016).
\newblock {XGB}oost: A scalable tree boosting system.

\bibitem[Deshpande and Jensen, 2016]{deshpande_jensen_2016}
Deshpande, S.~K. and Jensen, S.~T. (2016).
\newblock Estimating an {NBA} player’s impact on his team’s chances of
  winning.
\newblock {\em Journal of Quantitative Analysis in Sports}, 12(2):51--72.

\bibitem[Jain, 2022]{jain_2022}
Jain, A. (2022).
\newblock Complete guide to parameter tuning in {XGB}oost.

\bibitem[James et~al., 2013]{James2013}
James, G., Witten, D., Hastie, T., and Tibshirani, R. (2013).
\newblock {\em An Introduction to Statistical Learning: with Applications in
  R}.
\newblock Springer.

\bibitem[Kubatko et~al., 2007]{Kubatko_etal_2007}
Kubatko, J., Oliver, D., Pelton, K., and Rosenbaum, D.~T. (2007).
\newblock A starting point for analyzing basketball statistics.
\newblock {\em Journal of Quantitative Analysis in Sports}, 3(3).

\bibitem[Lindsey, 1963]{lindsey_1963}
Lindsey, G.~R. (1963).
\newblock An investigation of strategies in baseball.
\newblock {\em Operations Research}, 11(4):477–501.

\bibitem[Lock and Nettleton, 2014]{lock_nettleton_2014}
Lock, D. and Nettleton, D. (2014).
\newblock Using random forests to estimate win probability before each play of
  an {NFL} game.
\newblock {\em Journal of Quantitative Analysis in Sports}, 10(2):197--205.

\bibitem[Maddox et~al., 2022a]{maddoxetal_cbb_2022}
Maddox, J.~T., Sides, R., and Harvill, J.~L. (2022a).
\newblock Bayesian estimation of in-game home team win probability for college
  basketball.
\newblock {\em Journal of Quantitative Analysis in Sports}.

\bibitem[Maddox et~al., 2022b]{maddoxetal_nba_2022}
Maddox, J.~T., Sides, R., and Harvill, J.~L. (2022b).
\newblock Bayesian estimation of in-game home team win probability for national
  basketball association games.
\newblock {\em Journal of Sport Analytics}.

\bibitem[Pomeroy, 2012]{pomeroy_2012}
Pomeroy, K. (2012).
\newblock Ratings glossary: The kenpom.com blog.

\bibitem[{Pro Football Reference}, 2012]{sports_reference_2012}
{Pro Football Reference} (2012).
\newblock The p-f-r win probability model.
\newblock
  \url{https://www.sports-reference.com/blog/2012/03/features-expected-points/}.

\bibitem[{R Core Team}, 2021]{R}
{R Core Team} (2021).
\newblock {\em R: A Language and Environment for Statistical Computing}.
\newblock R Foundation for Statistical Computing, Vienna, Austria.

\bibitem[Ruscio and Brady, 2021]{ruscio_brady_2021}
Ruscio, J. and Brady, K. (2021).
\newblock Estimating win probability for nfl games.

\bibitem[Ryall, 2011]{ryall-2011}
Ryall, R. (2011).
\newblock {\em Predicting Outcomes in Australian Rules Football}.
\newblock PhD thesis, Royal Melbourne Institute of Technology University.

\bibitem[{Santos-Fernandez} et~al., 2019]{fernandez_2019}
{Santos-Fernandez}, E., Wu, P., and Mengersen, K.~L. (2019).
\newblock Bayesian statistics meets sports: a comprehensive review.
\newblock {\em Journal of Quantitative Analysis in Sports}, 15(4):289--312.

\bibitem[Terner and Franks, 2021]{terner_franks_2020}
Terner, Z. and Franks, A. (2021).
\newblock Modeling player and team performance in basketball.
\newblock {\em Annual Review of Statistics and Its Applications}, 8(1):1--23.

\bibitem[Troch, 2016]{troch_2016}
Troch, J. (2016).
\newblock A new look at college football tempo.

\bibitem[Wickham, 2021]{rvest}
Wickham, H. (2021).
\newblock {\em rvest: Easily Harvest (Scrape) Web Pages}.
\newblock R package version 1.0.2.

\end{thebibliography}

\end{document}